# Hyperbolic Bloch points in ferrimagnetic exchange spring

Javier Hermosa-Muñoz[1,2], Aurelio Hierro-Rodríguez[1,2*], Andrea Sorrentino[3], José I. Martín[1,2], Luis M. Alvarez-Prado[1,2], Eva Pereiro[3], Carlos Quirós[1,2], María Vélez[1,2*] and Salvador Ferrer[3,*]

[1]Depto. Física, Universidad de Oviedo, 33007 Oviedo, Spain.

[2]CINN (CSIC – Universidad de Oviedo), 33940 El Entrego, Spain.

[3]ALBA Synchrotron, 08290 Cerdanyola del Vallès, Spain.

*Email:* hierroaurelio@uniovi.es; mvelez@uniovi.es; ferrer@cells.es

**Abstract.** Bloch points in magnetic materials are attractive entities in view of magnetic information transport. Here, Bloch point configuration has been investigated and experimentally determined in a magnetic trilayer ($Gd_{12}Co_{88}/Nd_{17}Co_{83}/Gd_{24}Co_{76}$) with carefully adjusted composition within the ferrimagnetic $Gd_xCo_{1-x}$ alloys in order to engineer saturation magnetization, exchange length, and interlayer couplings (ferromagnetic *vs* antiferromagnetic). X-ray vector magnetic tomography has allowed us to determine experimentally Bloch point polarity (related to topological charge) and Bloch point helicity $\gamma$ (determined by magnetostatic energy). At the bottom layer (close to the ferromagnetic interface), Bloch points adopt a standard circulating configuration with helicity $\gamma$ close to $\pi/2$. Within the top layer (with much lower saturation magnetization), Bloch points nucleate within a Neel-like exchange spring domain wall created by the antiferromagnetic coupling and adopt an uncommon hyperbolic configuration, characterized by much larger helicity angles. Our results indicate a path for Bloch point engineering in future applications adjusting material parameters and domain wall characteristics.

Magnetic multilayers and patterned magnetic structures provide a rich playground for the design of magnetic textures with tailored topological characteristics beyond the limitations of single magnetic materials. For example, coupling between the layers in a synthetic antiferromagnet has allowed to stabilize meron and bimeron pairs[1] and novel vortex textures have been observed in magnetic nanocaps thanks to the competition between local/nonlocal interaction terms.[2] Bloch points (BPs) are attracting interest as magnetic point-singularities, with integer topological charge ($Q$),[3,4] since they mediate in many different reversal processes and topological transformations such as in magnetic nanospheres,[5] nanodisks,[6] domain wall (DW) cores in nanowires[7,8] and microstructures,[9] 3D magnetic metalattices,[10] *etc*. In extended samples, BPs usually appear within DW cores such as in periodic stripe domain patterns,[11] within vortex and skyrmion tubes in multilayers[12-15] and in other complex topological textures.[16]

Both theoretical and micromagnetic calculations predict very different BP configurations depending on the competition of exchange and magnetostatics in the surrounding regions.[4, 5, 17-22] In nanospheres, magnetostatic interactions appear as the most relevant term to determine the BP equilibrium configuration resulting in a preference of twisted BPs over radial hedgehogs at remanence.[5,19] Lower symmetry BPs have also been predicted in the presence of non-homogeneous magnetic fields,[21] with a natural length scale for spatial variations given by the material exchange length ($l_{ex}$).[22] BPs in nanowires adopt different configurations (ranging from circulating to hedgehog-like) depending on the interplay between exchange and magnetostatics as a function of wire radius and material parameters such as saturation magnetization ($M_S$) and anisotropy ($K$).[8,23-25] Also, it has been proposed that tailored exchange interactions in multilayers could provide a tool to manipulate BP configuration: *e.g.* at the interface of FeGe layers of different chirality[26,27] or at the interface of hard/soft exchange springs.[28]

On the other hand, the experimental characterization of BPs has only become possible in recent years thanks to the development of high-resolution X-ray magnetic vector imaging techniques.[11,29-30] However, up to now, most work has been focused on the determination of the basic topological properties of the singularities, such as topological charge, vorticity and emergent fields[9,11,29-30] and, a detailed experimental study of BP configuration as a function of their magnetic environment is still lacking.

Herein, we have prepared a $Gd_{12}Co_{88}/Nd_{17}Co_{83}/Gd_{24}Co_{76}$ trilayer[31] with engineered saturation magnetization (to tune magnetostatic energy and $l_{ex}$ in each layer) and interlayer couplings (to create a high DW density that favors BP nucleation). Then, X-ray magnetic vector tomography (XMVT) has been used to obtain a complete characterization of individual BPs in the sample, including polarity, helicity and topological charge in order to correlate them with material properties and DW configuration. A vertical segregation of BPs is observed with a clear dependence of BP helicity on saturation magnetization. At the high $M_S$ $Gd_{12}Co_{88}$ layer, the usual circulating BPs appear within Bloch DW cores, whereas, at the low $M_S$ $Gd_{24}Co_{76}$ layer, non-standard hyperbolic BPs are stabilized within an exchange spring domain wall (ESDW).

**RESULTS AND DISCUSSION**

*Theoretical description of BPs*

BPs are singular points in the magnetization, characterized by the condition $m_x = m_y = m_z = 0$.[32] For an axially symmetric BP, the magnetization unit vector $\mathbf{m} = \mathbf{M}(\mathbf{r})/M_S$ can be described with the ansatz proposed in refs. [5,18-22]

$$\mathbf{m} = (m_x, m_y, m_z) = (\sin\Theta\cos\Phi, \sin\Theta\sin\Phi, \cos\Theta) \quad (1)$$

Where the polar and azimuthal angles vary as

$$\Theta(\mathbf{r}) = p\theta + \frac{\pi}{2}(1-p) \quad (2)$$

$$\Phi(\mathbf{r}) = q\varphi + \gamma \quad (3)$$

in terms of the standard spherical coordinates $r, \theta$ and $\varphi$. BP configuration is determined by the parameters: polarity ($p$), vorticity ($q$) and helicity ($\gamma$). The topological monopole ($Q$) at the BP singularity is given by the combination of vorticity and polarity as $Q = pq$.[33] For example, for $q = 1$ and $p = +1$, BP configuration is a vortex at the equatorial plane with a tail-to-tail (T2T) DW at the symmetry axis (head-to-head (H2H) DW for $p = -1$).

Figure 1 shows several analytical BP models with $q = +1$, $p = +1$ and different helicity angles $\gamma$ between 0 and $\pi$. $\gamma = 0$ corresponds to the radial hedgehog with $\mathbf{m}$ pointing outwards along $\mathbf{r}$ at all angles (Fig. 1(a)). $\gamma = \pi/2$ is a circulating BP, characterized by a circulating vortex at the equator (Fig. 1(c)). $\gamma = \pi$ corresponds to the hyperbolic hedgehog (Fig. 1(e)): the magnetization points radially inwards at the equatorial plane and outwards at the north/south poles resulting in a hyperbolic configuration in the $x - z$ plane. Intermediate helicity values result in either twisted radial BPs for $0 < \gamma < \pi/2$ or twisted hyperbolic BPs ($\pi/2 < \gamma < \pi$) configurations (see Figs. 1(b)&(d) for $\gamma = \pi/4$ and $3\pi/4$, respectively). Equivalent configurations are found for negative polarity $p = -1$, only reversing the helicity ranges (twisted radial for $p = -1$ & $\pi/2 < \gamma < \pi$ and twisted hyperbolic for $p = -1$ & $0 < \gamma < \pi/2$).

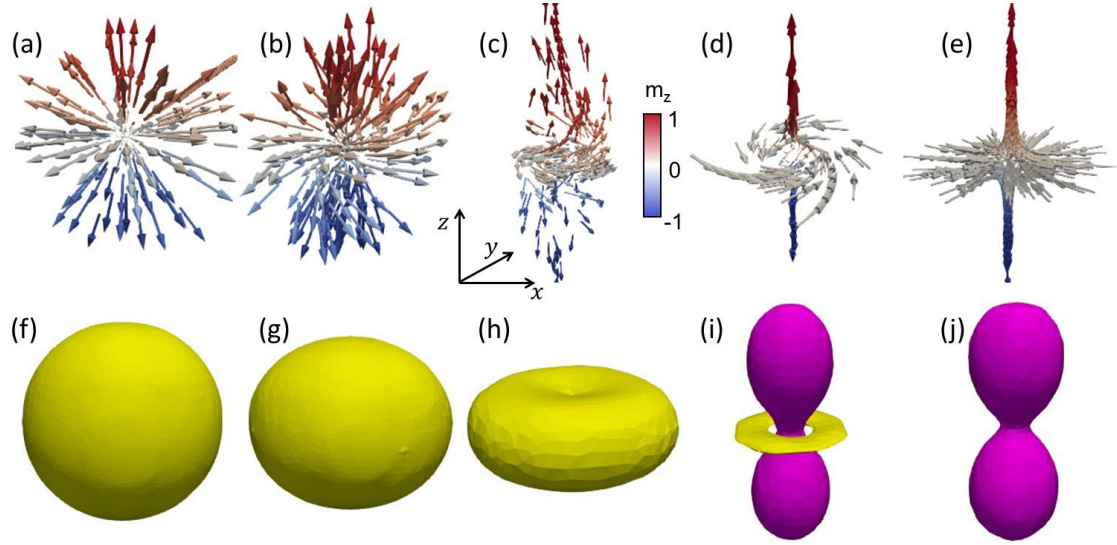

**Figure 1: Analytical model of symmetric Bloch points with different helicity $\gamma$.** Magnetic configuration of BPs with $q = +1$, $p = +1$ and (a) $\gamma = 0$; (b) $\gamma = \pi/4$; (c) $\gamma = \pi/2$; (d) $\gamma = 3\pi/4$; (e) $\gamma = \pi$. Constant divergence equisurfaces for (f) $\gamma = 0$, $\nabla \cdot \mathbf{m} = +0.025\ nm^{-1}$; (g) $\gamma = \pi/4$; $\nabla \cdot \mathbf{m} = +0.025\ nm^{-1}$ (h) $\gamma = \pi/2$, $\nabla \cdot \mathbf{m} = +0.025\ nm^{-1}$; (i) $\gamma = 3\pi/4$, $\nabla \cdot \mathbf{m} = +0.01\ nm^{-1}$ and $\nabla \cdot \mathbf{m} = -0.025\ nm^{-1}$; (j) $\gamma = \pi$, $\nabla \cdot \mathbf{m} = -0.025\ nm^{-1}$. Positive/Negative $\nabla \cdot \mathbf{m}$ equi-surfaces are indicated in yellow/purple, respectively.

Theoretical BP models show that the most relevant energy term to determine the equilibrium BP configuration is magnetostatics, which is directly linked to the helicity angle.[5,18-22] The relationship between BP helicity and magnetostatic energy can be visualized in terms of the magnetostatic charge surrounding the BP, that is proportional to the divergence of the magnetization. For $q = +1$, $\nabla \cdot \mathbf{m}$ is given by

$$\nabla \cdot \mathbf{m} = \frac{1}{r}[p \sin^2 \theta + \cos \gamma\, (1 + \cos^2 \theta)] \qquad (4).$$

Considering $p = 1$ (as in Fig. 1), $\nabla \cdot \mathbf{m}$ is positive at all angles for BPs with helicity in the range $0 \leq \gamma \leq \pi/2$. For the radial hedgehog ($\gamma = 0$), $\nabla \cdot \mathbf{m} = 2/r$ so that $\nabla \cdot \mathbf{m}$ equisurfaces are spheres centered at the singularity (see Fig. 1(f)). For circulating BPs with $\gamma = \frac{\pi}{2}$, $\nabla \cdot \mathbf{m} = \sin^2 \theta /r$ so that $\nabla \cdot \mathbf{m}$ equisurfaces take a torus shape (Fig. 1(h)). For $\gamma$ in the range $\frac{\pi}{2} < \gamma < \pi$ (twisted-hyperbolic BPs), positive $\nabla \cdot m$ regions at the BP equatorial plane coexist with a negative $\nabla \cdot m$ branch along the symmetry axis (Fig. 1(i)). Finally, at $\gamma = \pi$, the pure hyperbolic hedgehog, $\nabla \cdot \mathbf{m} = -\frac{2}{r}\cos^2 \theta$, so that $\nabla \cdot \mathbf{m}$ is negative at all angles (Fig. 1(j)).

Eq. (4) implies that, at the symmetry axis, $\nabla \cdot \mathbf{m}(\theta = 0, \pi)$ is independent of polarity and follows a simple $1/r$ dependence, so that helicity of the different BP configurations can be estimated as

$$\cos \gamma = r\, \nabla \cdot \mathbf{m}(\theta = 0, \pi)/2 \qquad (5).$$

For BPs in nanospheres, maxima in the magnetostatic energy correspond to the radial configurations $\gamma = 0, \pi$, and a minimum is found close to $\gamma = 0.6\,\pi$ depending on boundary conditions and model details.[5,18-22]. More detailed calculations indicate that equations (1-3) provide a good approximation only close to the BP core[4,21-22] (within distances of the order of the exchange length $l_{ex}$). Further away from the singularity, $\Theta(\mathbf{r})$ and $\Phi(\mathbf{r})$ deviate from this ideal configuration due to long range magnetostatic interactions with the environment.[22]

### *Domain walls in exchange coupled Gd₁₂Co₈₈/Nd₁₇Co₈₃/Gd₂₄Co₇₆ trilayer*

BP configuration has been studied within an 80 nm Gd₁₂Co₈₈/80 nm Nd₁₇Co₈₃/80 nm Gd₂₄Co₇₆ trilayer by a combination of XMVT[11] and micromagnetic simulations with Mumax³.[34] The sample was fabricated by magnetron sputtering,[31,35] choosing material parameters and interlayer couplings in order to create a variable magnetic environment at the different layers and induce a high density of DWs both parallel and perpendicular to the sample plane, as observed by XMVT in ref. [31]. This rich DW structure provides preferred loci for BP nucleation and determines the long-range magnetic environment of each singularity.

As sketched in Figs. 2(a-b), the central layer of the sample is made of the ferromagnetic Nd₁₇Co₈₃ alloy with saturation magnetization $M_S(NdCo) = 7 \times 10^5\ A/m$ and perpendicular magnetic anisotropy ($K_N = 10^5\ J/m^3$).[11] Its role is to create a pattern of up/down stripe domains, separated by Bloch walls,[11,32] that controls the magnetic configuration of the outer Gd$_x$Co$_{1-x}$ layers via exchange and magnetostatic interactions.

Top/Bottom layers are made of ferrimagnetic Gd$_x$Co$_{1-x}$ alloys in which Gd and Co normalized moments ($\mathbf{m_{Co}}$ and $\mathbf{m_{Gd}}$) are collinear and antiparallel[35,36] (*i. e.* $\mathbf{m_{Co}} = -\mathbf{m_{Gd}}$). The net magnetization of the Gd$_x$Co$_{1-x}$ alloy is given by $\mathbf{M}(GdCo) = M_{Co}\mathbf{m_{Co}} + M_{Gd}\mathbf{m_{Gd}} = (M_{Co} - M_{Gd})\mathbf{m_{Co}}$ with $M_{Co}$ and $M_{Gd}$ the saturation magnetization of Co and Gd ion sublattices, respectively. Then, the net unit magnetic moment of the alloy $\mathbf{m_{net}}(GdCo) = \frac{\mathbf{M}(GdCo)}{|M_{Co} - M_{Gd}|}$ will be aligned either with $\mathbf{m_{Co}}$ or $\mathbf{m_{Gd}}$ depending on the sign of $M_{Co} - M_{Gd}$. Gd$_x$Co$_{1-x}$/Nd₁₇Co₈₃ exchange at the interfaces is dominated by the ferromagnetic interaction between Co-Co moments.[36] Thus, the effective coupling between the net magnetic moments at both sides of each interface can be switched from parallel to antiparallel tuning the sign of $M_{Co} - M_{Gd}$ (see sketches in Figs. 2(a-b)).

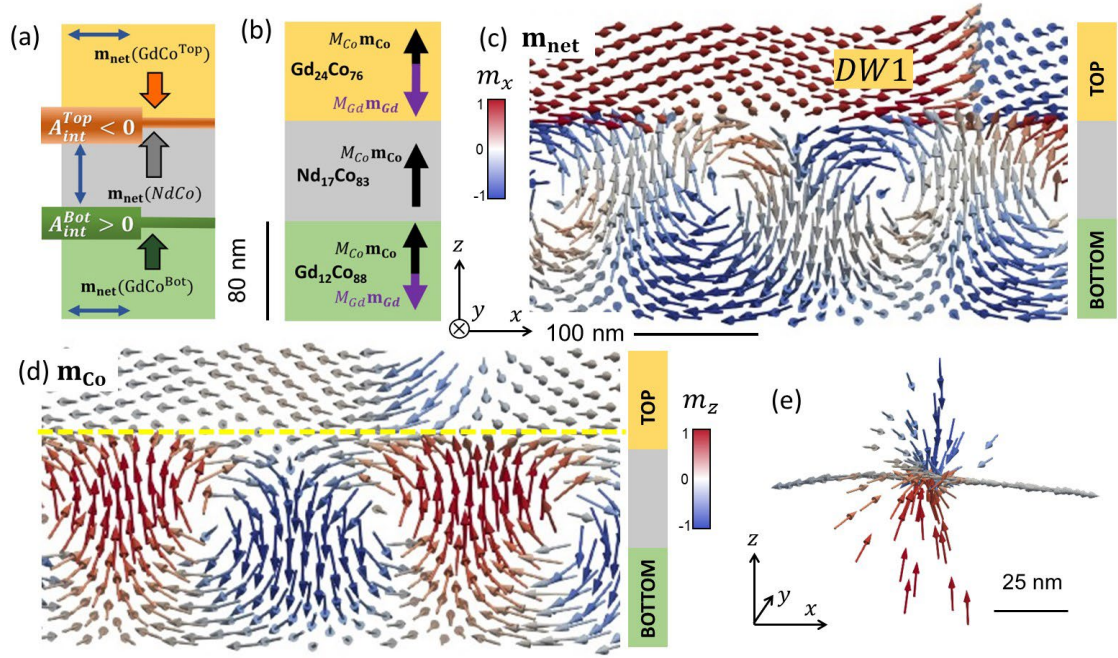

**Figure 2: Micromagnetic configuration of $Gd_{12}Co_{88}/Nd_{17}Co_{83}/Gd_{24}Co_{76}$ trilayer with an exchange spring domain wall**: Sketch of $Gd_{12}Co_{88}/Nd_{17}Co_{83}/Gd_{24}Co_{76}$ trilayer structure indicating (a) net magnetic moment in each layer and (b) individual magnetic moments of Co and Gd ions in the ferrimagnetic $Gd_xCo_{1-x}$ alloys. Note that the parallel alignment of Co moments throughout the sample results in an effective antiparallel coupling at the top $Nd_{17}Co_{83}/Gd_{24}Co_{76}$ interface, that is the origin of the Exchange Spring in the trilayer. Double arrows indicate easy anisotropy axis in each layer. (c) Cross section of micromagnetic simulation of $Gd_{12}Co_{88}/Nd_{17}Co_{83}/Gd_{24}Co_{76}$ trilayer at remanence in terms of net magnetic moment $\mathbf{m_{net}}$. A quasi-domain boundary within $GdCo^{Top}$ is marked as DW1. (d) Same micromagnetic simulation as in (c) in terms of Co magnetic moment $\mathbf{m_{Co}}$. The approximate location of the ESDW (defined by $m_z = 0$) is indicated by a dashed yellow line. (e) Detail of micromagnetic configuration around a BP nucleated at the intersection between the ESDW and a quasi-domain boundary within $GdCo^{Top}$.

The composition of the bottom $Gd_{12}Co_{88}$ layer ($GdCo^{Bot}$) is chosen so that $M_{Co}^{Bot} > M_{Gd}^{Bot}$.[35] Then, at the $GdCo^{Bot}/Nd_{17}Co_{83}$ interface the effective exchange $A_{int}^{Bot}$ between net magnetic moments $\mathbf{m_{net}}(GdCo^{Bot})$ and $\mathbf{m_{net}}(NdCo)$ is ferromagnetic since both of them are parallel to $\mathbf{m_{Co}}$ within their respective layers. On the contrary, the composition of the top $Gd_{24}Co_{76}$ layer ($GdCo^{Top}$) is selected so that $M_{Co}^{Top} < M_{Gd}^{Top}$ and $\mathbf{m_{net}}(GdCo^{Top}) = -\mathbf{m_{Co}^{Top}}$ (see sketch in Figs. 2(a-b)).[35] Then, at $Nd_{17}Co_{83}/GdCo^{Top}$ interface, the effective exchange $A_{int}^{Top}$ is negative since the ferromagnetic alignment of Co moments at both sides of the interface implies an antiparallel configuration of $\mathbf{m_{net}}(GdCo^{Top})$ and $\mathbf{m_{net}}(NdCo)$.

The consequence of the opposite signs for $A_{int}^{Bot}$ and $A_{int}^{Top}$ is the nucleation of a high density of DWs, as observed by XMVT[31] and micromagnetic simulations (Figs. 2(c-d)): the simulated magnetic configuration of the trilayer is shown in Fig. 2(c), with the standard $\mathbf{m_{net}}$ representation; in addition, Fig. 2(d) shows the equivalent $\mathbf{m_{Co}}$ representation, derived from the data in Fig. 2(c) with the rule $\mathbf{m_{Co}} = +\mathbf{m_{net}}$ at $GdCo^{Bot}$ and $NdCo$ layers and $\mathbf{m_{Co}} = -\mathbf{m_{net}}$ at $GdCo^{Top}$ in order to analyze the exchange spring created by Co moments.

At $GdCo^{Bot}$/Nd₁₇Co₈₃ interface, the out-of-plane domains created by $K_N$ at the Nd₁₇Co₈₃ layer are directly imprinted on $GdCo^{Bot}$ by the cooperative effect of exchange and magnetostatic interactions. The result is a periodic pattern of up/down stripe domains (with period 215 nm), separated by Bloch DWs with their corresponding closure domain structure, that fills the central and bottom parts of the sample (Nd₁₇Co₈₃ and $GdCo^{Bot}$ layers).

At Nd₁₇Co₈₃/$GdCo^{Top}$ interface, the competition between exchange interactions (that try to align $\mathbf{m_{Co}}$ at both sides of the interface) and magnetostatic interactions (that try to conserve the perpendicular component of $\mathbf{m_{net}}$ at both sides of the interface to minimize magnetostatic charges), creates an interfacial exchange spring that reverses the sign of $m_z$ by a gradual Néel-like rotation of $\mathbf{m_{Co}}$ across the trilayer thickness. The corresponding $m_z = 0$ surface is an ESDW that lies, approximately, 25 nm above the Nd₁₇Co₈₃/$GdCo^{Top}$ interface parallel to the sample plane (see dashed yellow line in Fig. 2(d)).

At the upper part of the trilayer, above the ESDW, the micromagnetic simulation shows large in-plane "quasi-domains",[31,32] *i.e.* regions of quasi-uniform $\mathbf{m_{net}}(GdCo^{Top})$, with orientation fixed by the antiparallel exchange with the closure domains of the Nd₁₇Co₈₃ stripe pattern and in-plane anisotropy of the $GdCo^{Top}$ layer, separated by quasi-domain boundaries such as DW1 in Fig. 2(c).

At the intersections between these different DWs present in the sample, it is possible to observe BP singularities whenever the condition $m_x = m_y = m_z = 0$ is fulfilled. For example, Fig. 2(e) shows a detail of the simulated micromagnetic configuration of a BP that is located at the intersection between the ESDW and a quasi-domain boundary.

Finally, an important feature in the trilayer design is the selection $M_S$ values at the different layers in order to tune the exchange/magnetostatic energy balance and the size of the relevant magnetic textures. At $GdCo^{Bot}$, Gd₁₂Co₈₈ composition corresponds to a saturation magnetization[35] $M_S(GdCo^{Bot}) = 5.1 \times 10^5 \, A/m$ with exchange length $l_{ex}^{Bot} = \sqrt{A/\frac{1}{2}\mu_0 M_S^2} \approx 7 \, nm$ (estimated with $A \approx 0.7 \times 10^{-11} \, J/m$ typical of amorphous Rare Earth-Transition Metal alloys[36,37]). At $GdCo^{Top}$, Gd₂₄Co₇₆ composition provides a lower saturation magnetization,[35] $M_S(GdCo^{Top}) = 8.9 \times 10^4 \, A/m$ which corresponds to a much larger exchange length $l_{ex}^{Top} = 38 \, nm$. As a comparison, the Néel-to-Bloch critical thickness for a layer with in plane anisotropy (such as $GdCo^{Top}$) can be estimated as[32] $t_c \approx 12 \, l_{ex}^{Top} = 450 \, nm$. This is well above the 80 nm $GdCo^{Top}$ thickness implying that Néel-like rotations will be favoured here. Also, $l_{ex}^{Top}$ is the natural length scale that sets the width of the ESDW across the thickness, the width of quasi-domain boundaries and, correspondingly the size of the inhomogeneous magnetic configuration associated to the BPs that nucleate within them.

*Experimental characterization of BPs by X-ray magnetic vector tomography in $Gd_{12}Co_{88}/Nd_{17}Co_{83}/Gd_{24}Co_{76}$ trilayer*

Figure 3(a) shows a Magnetic Transmission X-ray microscopy (MTXM) image of the $Gd_{12}Co_{88}/Nd_{17}Co_{83}/Gd_{24}Co_{76}$ trilayer taken with the full field microscope at the MISTRAL beamline of the ALBA synchrotron.[38]. The image has been measured at the Gd $M_{4,5}$ absorption edges with an oblique angle of incidence ($\theta = 35°$) so that magnetic contrast is sensitive to the magnetization components $m_x$ and $m_z$ of Gd ions[31] at top and bottom $Gd_xCo_{1-x}$ layers. The detailed 3D magnetization configuration of the trilayer was obtained from the tomographic reconstruction of the changes in magnetic contrast of over 150 MTXM images with variable angle of incidence and in-plane sample orientation with a dedicated reconstruction algorithm.[39] As reported in ref. [31], the short scale pattern of clear/dark lines with period 215 nm corresponds to the out-of-plane oscillation of the magnetization in the stripe domains in the lower part of the trilayer. Clear/dark regions extending over several stripe periods correspond to quasi-domains at $GdCo^{Top}$ with different in-plane magnetization orientation (indicated by yellow arrows in Fig. 3(a)).

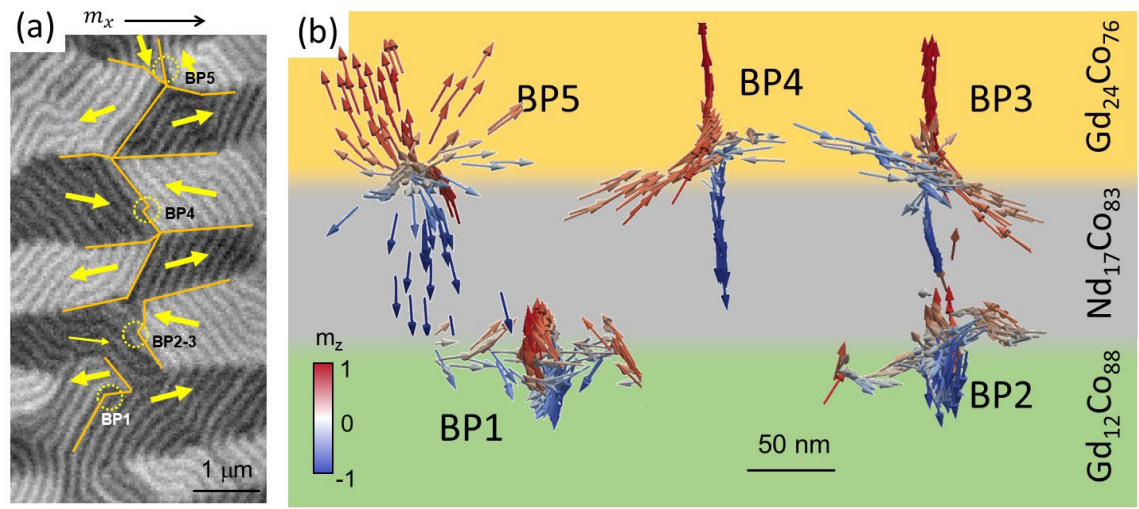

**Figure 3. Observation of Bloch Points by X-ray Magnetic Vector Tomography at different vertical positions in the trilayer.** (a) MTXM image at 35° X-ray angle of incidence with both $m_x - m_z$ contrast. Thin orange solid lines mark DWs between quasi-domains. Yellow arrows indicate average magnetization orientation at each quasi-domain. Dashed circles indicate selected BP locations. b) Detail of magnetic configuration around selected BPs at different vertical positions within the trilayer. Note that circulating BPs (BP1, BP2) appear close to the bottom $Gd_{12}Co_{88}/Nd_{17}Co_{83}$ interface with parallel coupling whereas hyperbolic BPs (BP3, BP4, BP5) tend to nucleate near the top $Nd_{17}Co_{83}/Gd_{24}Co_{76}$ interface.

24 BPs have been identified within the 8 $\mu$m × 8 $\mu$m × 300 nm volume of the magnetic tomogram with the condition $m_x = m_y = m_z = 0$ and topological charge $|Q| \approx 1$. Figure 3(b) shows the magnetic configuration $\mathbf{m}(\mathbf{r})$ in the vicinity of several BPs found at different vertical positions of the trilayer at the locations marked by dashed circles in Fig. 3(a). The observed BPs tend to appear at quasi-domain boundaries (marked by thin orange lines in Fig. 3(a)) either near the $GdCo^{Bot}/NdCo$ interface (*e. g.* BP1 and BP2 in Fig. 3(b)) or at the ESDW across the thickness within the $GdCo^{Top}$ layer (*e. g.* BPs 3-5 in Fig. 3(b)). At $GdCo^{Bot}$, BP1 and BP2 display a vortex like circulation of the magnetization around a horizontal symmetry axis hosting a linear DW, H2H in this case. At $GdCo^{Top}$, BPs display less symmetric $\mathbf{m}(\mathbf{r})$ configurations with characteristic hyperbolic cross sections (see *e. g.* BP3 and BP4) and, in certain cases, a more radial magnetization arrangement at their equatorial plane (see *e.g.* BP5).

Basic parameters of each BP configuration such as polarity, vorticity, helicity and topological charge have been estimated from the combined information of magnetization divergence $\nabla \cdot \mathbf{m}$ and topological charges computed from experimental $\mathbf{m}(\mathbf{r})$ data. The process is detailed below, and the results are summarized in Table 1:

i) First, the approximate symmetry axis of $\mathbf{m}(\mathbf{r})$ in the vicinity of each singularity is derived from the shape of magnetic divergence equi-surfaces, as shown in the magnetic tomogram of Fig. 4(a) for BP4 and $\nabla \cdot \mathbf{m} = -0.03$ nm$^{-1}$ at $GdCo^{Top}$. In this case, it presents an elongated shape with a central waist, qualitatively similar to the twisted hyperbolic BPs in Fig. 1(d), corresponding to a vertical symmetry axis.

ii) BP polarity is defined from the magnetization reversal sense along the symmetry axis: T2T for the singularity in Fig. 4(a), corresponding to $p = +1$.

iii) Topological charge $Q$ is computed from the flux of emergent field across a closed surface containing each singularity, following the procedure described in refs. [9,11], with $Q = +1$ for BP4.

iv) Vorticity is estimated from the comparison of experimental polarity and topological charge with the condition $Q = pq$. We obtain $q = +1$ in all cases.

| Layer | Polarity ($p$) | Topological charge ($Q$) | Vorticity ($q$) | Helicity ($\cos\gamma$) | | |
|---|---|---|---|---|---|---|
| | | | | $[-1,-0.2]$ | $(-0.2,0.2)$ | $[+0.2,+1]$ |
| $GdCo^{Top}$ | $-1$ (H2H) | $-1$ | $+1$ | | | 6 |
| | $+1$ (T2T) | $+1$ | $+1$ | 9 | | |
| $GdCo^{Bot}$ | $-1$ (H2H) | $-1$ | $+1$ | | 7 | |
| | $+1$ (T2T) | $+1$ | $+1$ | | 2 | |

**Table 1. Number of BPs and their characteristic parameters identified within the experimental tomogram of the Gd$_{12}$Co$_{88}$/Nd$_{17}$Co$_{83}$/Gd$_{24}$Co$_{76}$ trilayer.**

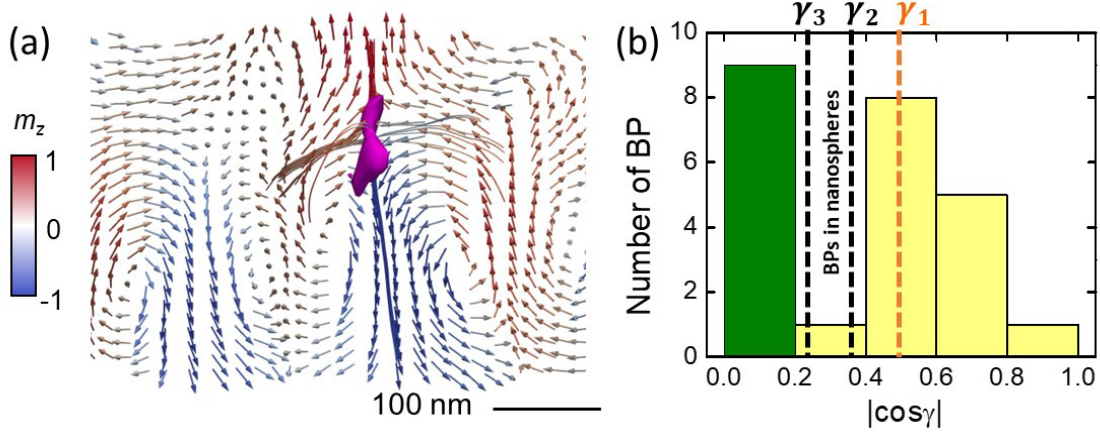

**Figure 4. Statistics of hyperbolic vs circulating Bloch points at $GdCo^{Top}$ and $GdCo^{Bot}$ within the magnetic tomogram.** a) Cross section of the magnetic tomogram $\mathbf{m}(\mathbf{r})$ showing BP4 within a domain wall between quasi-domains at $GdCo^{Top}$. Note the central T2T reversal along the symmetry axis of BP4 and the elongated shape of constant divergence equi-surface $\nabla \cdot \mathbf{m} = -0.03\ nm^{-1}$ indicated in purple, typical of twisted hyperbolic BPs. b) Histogram of the number of BPs with a given value of $|\cos\gamma|$: $GdCo^{Bot}$, green bar; $GdCo^{Top}$, yellow bars. $\gamma_1 - \gamma_3$ are theoretical predictions from analytical models.[5,18-22]

v) Finally, helicity $\gamma$ is estimated from the magnetic divergence at the BP symmetry axis using eq. (5). For not fully symmetric BPs, $\cos\gamma$ is calculated from the average between $\nabla \cdot \mathbf{m}(\theta = 0)$ and $\nabla \cdot \mathbf{m}(\theta = \pi)$ within 40-60 nm of the singularity (see Supporting Information for details). In particular, $\cos\gamma \approx -0.55 \pm 0.10$ is estimated for the BP4 in Fig. 4(a).

The statistics in Table 1 show two clear trends: first, an inverse correlation between the signs of polarity and $\cos\gamma$ at $GdCo^{Top}$ (note the empty diagonal in the top row of Table 1), that implies that all the observed BPs have a hyperbolic character; second, larger $|\cos\gamma|$ values at $GdCo^{Top}$ than at $GdCo^{Bot}$ with a threshold at $|\cos\gamma| = 0.2$.

These differences can be seen in more detail in the histogram of the number of BPs *vs.* $|\cos\gamma|$ at each $Gd_xCo_{1-x}$ layer shown in Fig. 4(b): at $GdCo^{Top}$, the number of BPs detected is maximum in the interval $0.4 < |\cos\gamma| < 0.6$ with an average value $\langle|\cos\gamma|\rangle_{Top} = 0.57$ whereas at $GdCo^{Bot}$ there is a more narrow distribution, with all detected BPs in the interval $|\cos\gamma| < 0.2$ and $\langle|\cos\gamma|\rangle_{Bot} = 0.15$. In terms of the theoretical BP classification shown in Fig. 1, these results imply a preference for twisted hyperbolic BPs at $GdCo^{Top}$ with helicity in the range $0.56\pi < \gamma < \pi$ for $p = +1$ ($0 < \gamma < 0.44\pi$ for $p = -1$) and a preference of configurations closer to the ideal circulating BP with $\gamma = \frac{\pi}{2}$ at $GdCo^{Bot}$ (*i.e.* with helicity in the $0.44\pi < \gamma < 0.56\pi$ range).

The observed preference for hyperbolic *vs.* circulating BPs can be ascribed to the different character of DWs in each layer. At $GdCo^{Bot}$, with a large $M_S$ and short $l_{ex}^{Bot} = 7\ nm$, BPs appear within Bloch walls in the stripe domain pattern. In this case, the presence of a BP singularity implies the sign reversal of in-plane magnetization along DW line, without a change in the surrounding closure vortex, very similar to BPs observed in Bloch DWs of ferromagnetic microstructures,[9] nanowires,[7,8] and in weak stripe domains.[11] In contrast, at $GdCo^{Top}$, with low $M_S$ and large $l_{ex}^{Top} = 38$ nm, BPs are hosted at the intersections between the Néel like ESDW and quasi-domain boundaries. In the case of BP4, shown in Fig. 4(a), it is located in between two regions with opposite in-plane magnetizations oriented almost head-to-head towards a quasi-domain boundary lying along the stripe pattern orientation. Here, the magnetization rotates out-of-plane, in an equivalent way to DW1 in the micromagnetic simulations of Fig.2(d). The vertical antiparallel orientation of magnetic moments at ESDW is stabilized by magnetostatic interactions due to the opposite signs of $\mathbf{m}_{Co}$ and $\mathbf{m}_{net}$ at $GdCo^{Top}$. This creates a hyperbolic configuration of the magnetization in the vertical cross section of the quasi-domain boundary, that is translated to the BP within it.

There are other situations in which magnetostatic terms have been shown to stabilize hyperbolic magnetic textures in confined geometries, as *e.g.* half-hedgehogs in nanocaps[40] or theoretical studies of BPs in nanospheres.[5,18-22] The optimum helicity values reported depend on the contributions to the magnetostatic energy considered in each model. One of the first predictions was based on the principle of pole avoidance, *i.e.* zero net magnetization divergence around the BP, with optimum helicity[18] $\gamma_1 = 2\pi/3$ ($cos\gamma_1 = -0.5$). Further refinement, considering stray and demagnetizing fields created by magnetic charges at the nanosphere boundary, resulted in smaller helicity values[19] $\gamma_2 = 0.62\pi$ ($cos\gamma_2 = -0.37$) and $\gamma_3 = 0.58\pi$ ($cos\gamma_3 = -0.25$).[5] More recently,[22] detailed calculations allowing a spatially variable helicity found an optimum range $\gamma(r) = 0.62\pi - 0.4\pi$ with $|\cos \gamma(r)| < 0.36$. These values are plotted as vertical dashed lines in Fig. 4(b) for comparison. In our case, BP statistics at $GdCo^{Top}$ show a preferred helicity value close to the predictions of the pole avoidance model[18] $cos\gamma_1 = -0.5$. This is consistent with the smaller relevance of stray fields for the singularities buried within an extended magnetic trilayer studied here in comparison with BPs confined in nanospheres.

**CONCLUSIONS**

The vector magnetization configuration $\mathbf{m}(\mathbf{r})$ of a Gd$_{12}$Co$_{88}$/Nd$_{17}$Co$_{83}$/Gd$_{24}$Co$_{76}$ trilayer, with tailored $M_S$ and interlayer couplings has been studied by XMVT and micromagnetic simulations. The systematic characterization of BPs in the magnetic tomogram in terms of their basic parameters (polarity, vorticity and helicity) reveals a clear dependence of BP configuration with the material parameters and domain wall characteristics within each layer.

At the bottom Gd$_{12}$Co$_{88}$ layer, BPs nucleate within the Bloch domain walls that separate up/down stripe domains, with a typical circulating configuration characterized by small values of $|cos\gamma|$ below 0.2, *i. e.* $\gamma \approx \frac{\pi}{2}$. On the other hand, at the top Gd$_{24}$Co$_{76}$ layer, with low $M_s$ and much larger exchange length, BPs appear at the intersection between the ESDW (created by antiferromagnetic coupling at the interface) and in-plane quasi-domain boundaries. In this case, BPs with twisted hyperbolic configurations are observed with much larger helicity angles. The average $\langle |\cos\gamma| \rangle_{Top} = 0.57$ is close to $\gamma \approx 2\pi/3$, corresponding to the estimates from the pole avoidance principle. Overall, this work shows how different magnetic configurations in heterogeneous systems give rise to different types of Bloch points which provides a path for determining specific Bloch points in view of applications.

**METHODS**

***Sample fabrication*** A 80 nm Gd$_{12}$Co$_{88}$/80 nm Nd$_{17}$Co$_{83}$/80 nm Gd$_{24}$Co$_{76}$ trilayer has been prepared by co sputtering on 50 nm thick Si-N membranes as reported before.[31,35] Vibrating Sample Magnetometry (VSM) and Transverse MOKE hysteresis loops measured on thin films with similar composition as each layer in the trilayer provide information regarding the saturation magnetization and the uniaxial anisotropy in each layer: $M_s(GdCo^{BOT}) = 5.1 \times 10^5 A\, m^{-1}$; $K(GdCo^{BOT}) = 5.1 \times 10^3 J\, m^{-3}$; $M_s(NdCo) = 7 \times 10^5 A\, m^{-1}$; $K(NdCo) = 10^5 J\, m^{-3}$; $M_s(GdCo^{TOP}) = 8.9 \times 10^4 A\, m^{-1}$; $K(GdCo^{TOP}) = 1.3 \times 10^4 A\, m^{-1}$. As the anisotropy of the central Nd$_{17}$Co$_{83}$ layer is an order of magnitude larger than in Gd$_x$Co$_{1-x}$ layers, its stripe domain pattern provides the pinned magnetic configuration within the trilayer. An out-of-plane demagnetizing cycle was applied to the trilayer in order to create a high DW density due to the competition of interactions and to favor BP nucleation within the sample.

***X-ray Magnetic vector Tomography*** The sample was mounted on the high precision rotary stage of the full field X-ray transmission microscope at the MISTRAL beamline of the ALBA Synchrotron.[38] One hundred nm diameter gold nanoparticles were sprinkled onto its surface to serve as fiducials for accurate projection alignment to a common rotation axis prior to the tomography reconstruction. The sample was illuminated with circularly polarized X-rays to exploit magnetic contrast from X-ray magnetic circular dichroism at the Gd M4 (1221.9 eV) and Gd M5 (1189.6 eV) absorption edges. The sample was rotated around an axis parallel to the sample surface to acquire a tilt series of images (at 2° intervals in the angular range $\theta = \pm 26°$ and at 1° intervals in the angular ranges $[-55°, -26°]$ and $[26°, 55°]$). As the tomographic reconstruction requires two orthogonal tilt series, the sample was rotated $\varphi \sim 90°$ (actually, 95°) in-plane before acquiring the second tilt series. A total of 150 angular images were acquired with exposition times around 5 s. For larger angles exposition times were increased to keep similar signal-to-noise ratios. Since the Gd M4 edge and the Gd M5 edge have opposite signs of the X-ray magnetic circular dichroic factor, images have been acquired at both edges at each $(\theta, \varphi)$ orientation. Adding or subtracting the logarithm of individual transmittance images at M4 and M5 edges allows us to obtain the charge (TXM) and magnetic (MTXM) contrast images.[31]

The reconstruction of the datasets is the key to obtain the 3D magnetic moment configuration in the trilayer. This reconstruction is performed using a vector tomography algorithm,[39] to recover the **m(r)** tomogram of normalized Gd magnetic moments with voxel size $(13.5\ nm)^3$, corresponding to the measured angular dependent magnetic contrast in the tilt series of images. The resolution of the resulting tomogram can be estimated as 30 nm lateral (given from the lateral resolution of the X-ray microscope) and 65 nm axial (estimated from the decay of the magnetic signal outside the sample along $z$ axis).[31]

*Micromagnetic simulations* The magnetic configuration of the sample at remanence after saturation in an out-of-plane magnetic field of 2T has been calculated with[34] Mumax³ with material parameters obtained from the experimental characterization of individual thin films with the same composition as the different layers in the measured trilayer.[35,37] Simulation size is 7.68 $\mu$m × 7.68 $\mu$m × 240 nm with a discretization of 5 nm. Saturation magnetization and magnetic anisotropy in each layer are the same as the values obtained experimentally, as listed above. Easy anisotropy axes are oriented along $x$ direction at top/bottom $Gd_xCo_{1-x}$ layers and along $z$ direction (out-of-plane) at the central $Nd_{17}Co_{83}$ layer. Exchange stiffness is $A_{NdCo} = 0.8 \times 10^{-11}\ J/m$ and $A_{GdCo} = 0.7 \times 10^{-11}\ J/m$ typical of Co-Co exchange in amorphous Rare Earth-Transition Metal alloys.[36-37] For magnetic moments at $Gd_xCo_{1-x}/Nd_{17}Co_{83}$ interfaces, an average $|A_{int}| = 0.75 \times 10^{-11}\ J/m$ is considered but with different signs in order to reflect the parallel/antiparallel alignment of **m**$_{net}$ and **m**$_{Co}$ at the different layers. In particular, $A_{int}^{Bot} = +0.75 \times 10^{-11}\ J/m$ since **m**$_{net}$ is parallel to **m**$_{Co}$ both at bottom and central layers. On the other hand, $A_{int}^{Top} = -0.75 \times 10^{-11}\ J/m$ is negative since **m**$_{net}$ is antiparallel to **m**$_{Co}$ at the top $Gd_{24}Co_{76}$ layer.

**Acknowledgements** The ALBA Synchrotron is funded by the Ministry of Research and Innovation of Spain, by the Generalitat de Catalunya and by European FEDER funds. This work has been supported by Spanish MCIN/AEI/10.13039/501100011033/FEDER,UE under grants

PID2019-104604RB and PID2022-136784NB and by Asturias FICYT under grant AYUD/2021/51185 with the support of FEDER funds.

**Supporting Information Available**

Description of Helicity estimation procedure for ideal, simulated and experimental BPs